\documentclass[twocolumn,amsmath,amssymb,aps]{revtex4-2}
\usepackage{lineno,hyperref}
\usepackage{bm}
\usepackage{amsmath}
\usepackage{gensymb}
\usepackage{epstopdf}
\usepackage{booktabs}

\usepackage{graphicx}
\usepackage{ulem}
\usepackage{xcolor}
\usepackage{natbib}
\usepackage{hyperref}
\usepackage{cleveref}

\begin{document}

\title {Rare observation of spin-gapless semiconducting characteristics and related band topology of quaternary Heusler alloy CoFeMnSn}

\author{Shuvankar Gupta$^{1}$}
\thanks{Contributed equally}
\author{Jyotirmoy Sau$^2$}
\thanks{Contributed equally}
\author{Manoranjan Kumar$^2$}
\email{manoranjan.kumar@bose.res.in}
\author{Chandan Mazumdar$^1$}
\email{chandan.mazumdar@saha.in}

\affiliation{$^1$Condensed Matter Physics Division, Saha Institute of Nuclear Physics,  A CI of Homi Bhabha National Institute, 1/AF, Bidhannagar, Kolkata 700064, India}
\affiliation{$^2$Department of Condensed Matter Physics and Material Science,
S. N. Bose National Centre for Basic Sciences, JD Block, Sector III, Salt Lake, Kolkata 700106, India}

\date{\today}

\begin{abstract}
 In this paper, we report the theoretical investigation and experimental realization of a new spin-gapless semiconductor (SGSs) compound CoFeMnSn belonging to the family of quaternary Heusler alloys. Through the use of several ground-state energy calculations, the most stable structure has been identified. Calculations of the spin-polarized band structure in optimized structure's reveals the  SGS nature of the compound. The compound form in an ordered crystal structure and exhibit a high ferromagnetic transition temperature (T$_{\rm C}$ = 560 K), making the material excellent for room temperature applications. Adherence of saturation magnetization to the Slater-Pauling rule, together with the nearly temperature-independent resistivity, conductivity, and carrier concentration of the compound in the temperature regime 5$-$300 K along with the low value of anomalous Hall conductivity (AHC) further confirms the SGS nature. Theoretical calculations also reveal the robustness of the SGS state due to lattice contraction and one can obtain a high value of intrinsic AHC using hole doping. Combined SGS and topological properties of the compound make CoFeMnSn suitable for spintronics and magneto-electronics devices.

\end{abstract}
\maketitle

\section{\label{sec:Introduction}Introduction}

Spin-polarized ferromagnetic materials with high Curie temperatures are an essential requirement in the emerging technology field of magneto-electronics and spintronics~\cite{katsnelson2008half,felser2007spintronics}. Half metallic ferromagnets (HMFs) are a special kind of such materials that can show 100 \% spin polarization due to its unique band structure in which one sub-band behaves like a metal while the other sub-band acts like a semiconductor~\cite{de1983new}. Recently, a fascinating new sub-class of compounds, known as spin gapless semiconductors (SGSs)~\cite{wang2008proposal}, belonging to such type of materials has been discovered and become the subject of intense investigation. In comparison to standard HMFs, one type of spin-polarized sub-band of SGSs has a band structure similar to that of a semiconductor, whereas the other sub-band has a zero band gap at the Fermi level instead of metallic behavior in the HMF. SGSs are not only able to produce 100 \% spin polarization in the presence of such a distinctive band structure, but they also display extremely high electrical mobility~\cite{wang2016recent,rani2020spin,yue2020spin}. Additionally, one can simply adjust and switch between \textit{n}- and \textit{p}-type spin-polarized carriers by applying external perturbations (such as an electric field, pressure, or magnetic field), making these materials the ideal choice for magneto-electronics/spintronics applications~\cite{bainsla2016equiatomic}.

The Heusler alloy type of compounds, having very favorable crystal chemistry, is generally considered to be quite conducive to realize materials that meet the stringent requirements for semiconductor spintronics~\cite{felser2007spintronics,tavares2022heusler}. In fact, the half-Heusler compound NiMnSb was the first compound to be proposed as HMF through band structure calculations~\cite{de1983new}. The interesting ferromagnetic character of Heusler alloys, which often are made up of non-ferromagnetic constituent elements, made them a scientific wonder when they were first discovered in 1905~\cite{graf2013magnetic}. Subsequently, two major types of Heusler alloys are considered: i) full Heusler alloy represented stoichiometrically by the formula X$_2$YZ ii) half Heusler alloy represented as XYZ, where X and Y are transition elements and Z is a \textit{sp}-group element~\cite{graf2013magnetic,graf2011simple}. Recently, another new variant quaternary Heusler represented by $XX'YZ$ has been introduced~\cite{dai2009new,bainsla2016equiatomic,alijani2011quaternary,alijani2011electronic,gupta2022coexisting,PhysRevB.108.045137,gupta20234d,gupta2023experimental}. As most of the practical usages of spintronics require room temperature applications, Co-based full/quaternary Heusler alloys have attracted a lot of attention due to their high Curie temperature (T$_{\rm C}$) and tunable electronic structure~\cite{graf2011simple,bombor2013half,rani2017structural,shigeta2018pressure,alijani2011electronic}. In spite of a large number of Heusler alloys have been theoretically proposed to possess SGS characteristics~\cite{xu2013new,gao2019high,gao2013antiferromagnetic}, in practice only handful numbers of materials have been such compound experimentally found so far. The development of this area of study is heavily dependent on the finding of new quaternary Heusler alloys, since only three of the compounds are experimentally realized compounds so far. In the full-Heusler alloy family, such compound experimentally found so far is on Mn$_2$CoAl\cite{ouardi2013realization}. Later, SGS nature was also confirmed in two other quaternary Heusler compounds CoFeMnSi~\cite{bainsla2015spin} and CoFeCrGa~\cite{bainsla2015origin}. Based on this very limited set of only the three above-mentioned compounds, it is very difficult to find any generalized rule that can help us in predicting new compounds having the potential to exhibit SGS characteristics. The only clue that one can find that Heusler alloys having valence electron counts (VEC) of 26 (Mn$_2$CoAl and CoFeCrGa) and 28 (CoFeMnSi)~\cite{wang2016recent} might have a better chance to be SGS.  It should also be highlighted that another novel variation, fully compensated ferrimagnetic SGS, has been identified in CrVTiAl (VEC = 18)~\cite{venkateswara2018competing}.

In the present work, we report a new quaternary SGS Heusler compound CoFeMnSn having VEC 28, investigated through comprehensive experimental and theoretical studies. SGS nature in CoFeMnSn has been confirmed through combined First principle theoretical calculations on the optimized structure, experimental validation of the Slater-Pauling rule, observation of temperature-independent resistivity/conductivity and carrier concentration, low value of anomalous hall conductivity (AHC) and linear magnetoresistance behaviour in the experimentally investigated regime 5$–$300 K. As SGS compounds are extremely sensitive to external factors (such as pressure, magnetic fields, and doping), we have also theoretically simulated the effect of pressure on its electronic structure. Our calculation suggests that CoFeMnSn changes from SGS to half-metal under the application of hydrostatic pressure, with an appearance of the minority-spin band gap, making the material still applicable for spintronic devices such as spin valves~\cite{ikhtiar2014magneto}, spin injectors~\cite{saito2013spin}, and magnetic tunnel junctions~\cite{kubota2009half}.

Another highly intriguing topics in condensed matter physics is the genesis of the Anomalous Hall effect (AHE) in magnetic materials, and it is still ambiguous whether AHE contributes in intrinsic or extrinsic way. In ferromagnetic materials, there is an anomalous term that is proportional to spontaneous magnetization in addition to the typical Hall effect that results from the Lorentz force deflecting moving charge carriers in a magnetic field. The AHE is caused by three mechanisms: skew scattering, side jump, and intrinsic deflection \cite{Nagaosa}. The Berry curvature of the occupied electronic Bloch states is closely connected to the intrinsic Karplus-Luttinger (K-L) mechanism \cite{Jungwirth, Gradhand_2012}. The AHE has long been considered the primary indicator of finite magnetization, in ferromagnetic materials. However, it has recently been discovered that the inherent value of AHE is often controlled by its non-vanishing Berry curvature rather than being directly related to magnetization \cite{Kubler2014}. Our theoretical calculation on CoFeMnSn reveals an interesting feature where hole doping would result in development of large Berry curvature, resulting a many-fold increase of intrinsic AHC.

\section{\label{sec:Methods}Methods}

\subsection{Experimental}
The poly-crystalline compound CoFeMnSn was synthesized by taking high-purity elements ($>$ 99.9 \%) through arc melting technique under an inert argon gas atmosphere. The prepared poly-crystalline sample was then annealed at 1073 K for 15 days in a vacuum followed by quenching in ice water. Room temperature powder X-ray diffraction (XRD) spectrum was taken using Cu-K$\alpha$ radiation  (TTRAX-III diffractometer, Rigaku, Japan). Full Rietveld refinement of the XRD data was done using FullProf software packages~\cite{rodriguez1993recent}. High-temperature VSM (Model EV9, MicroSense, LLC Corp., USA) was used to measure the magnetization (M) \textit{vs.} temperature (T) in the range of 300$–$700 K. M \textit{vs.} field (H) measurements were carried out using SQUID VSM (Quantum design Inc.(USA)) in the  magnetic fields up to $\pm$70 kOe at 5 K and 300 K. Resistivity ($\rho$(T)), magneto-resistance (MR), carrier concentration (\textit{n(T)}) and Hall measurements were performed using Physical Property Measurement System (Quantum design Inc. (USA)) utilizing the standard four-probe technique.

\subsection{Computational}

First-principles electronic structure calculations have been used to examine the band structures and transport characteristics of CoFeMnSn. The calculations were carried out using the projected augmented wave approach as implemented in the Vienna ab-initio simulation package (VASP) \cite{PE} to model the electronic structure of the system. The electronic exchange-correlation functional was approximated by the Perdew-Burke-Ernzerhof (PBE) \cite{burke} type generalized gradient approximation (GGA) \cite{perdew}. We have also started the DFT computation for this material by adding a coulomb repulsion. For the treatment of the exchange and correlation effects for Mn (3\textit{d}) electrons. For the plane-wave basis, a kinetic energy cutoff of 600 eV was used. The Brillouin zone (BZ) sampling utilized a $\Gamma$-centered 12$\times$ 12$\times$ 12 Monkhorst-pack \textit{k}-point mesh, and the electronic integral over the Brillouin Zone (BZ) was approximated using the Gaussian smearing method with a width of 0.05 eV. The electronic properties are calculated with and without spin-orbit coupling (SOC) as per requirement. The DFT band structure is fitted to calculate tight binding parameters using  a plane-wave basis state. The AHC and Berry curvature are estimated using the Wannier90 \cite{Pizzi_2020,marzari_97} and Wanniertools \cite{WU} codes. A dense 501 $\times$ 501 $\times$ 501 k-grid was used for the AHC calculations, with the tight-binding model's limit.

\section{Results and Discussion}

\subsection{\label{sec:DOS1}Structure optimization and electronic structure calculations}


\begin{figure}[ht]
\centerline{\includegraphics[width=.48\textwidth]{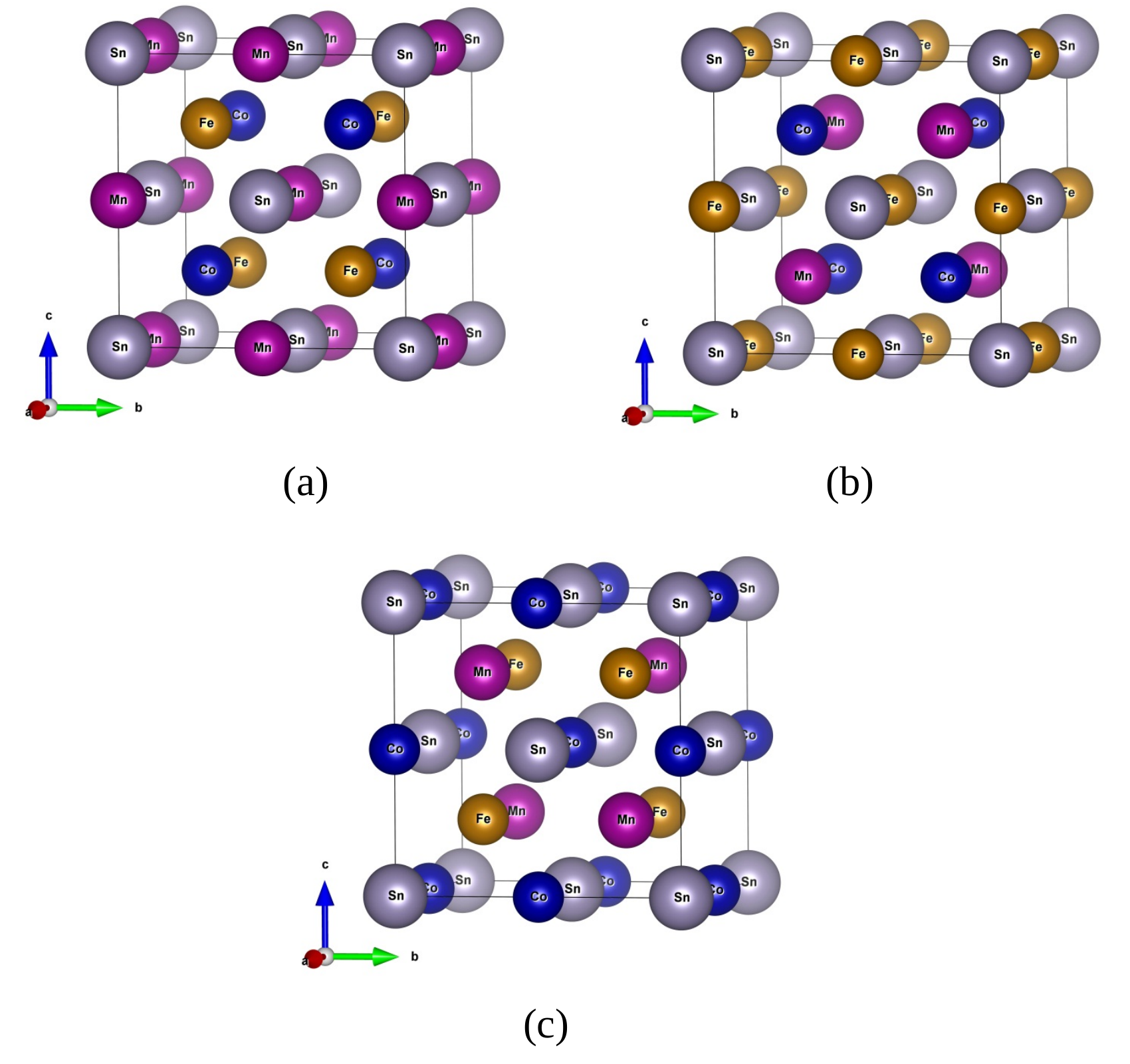}}
{\caption{Unit cell representation of (a) Type-1 (b) Type-2 (c) Type-3 ordered structure. Color representations of atoms, Co: blue ball, Fe: yellow ball, Mn: magenta ball and Sn: grey ball.}\label{Fig_2}}
\end{figure}

\begin{figure*}[t]
\centerline{\includegraphics[width=.96\textwidth]{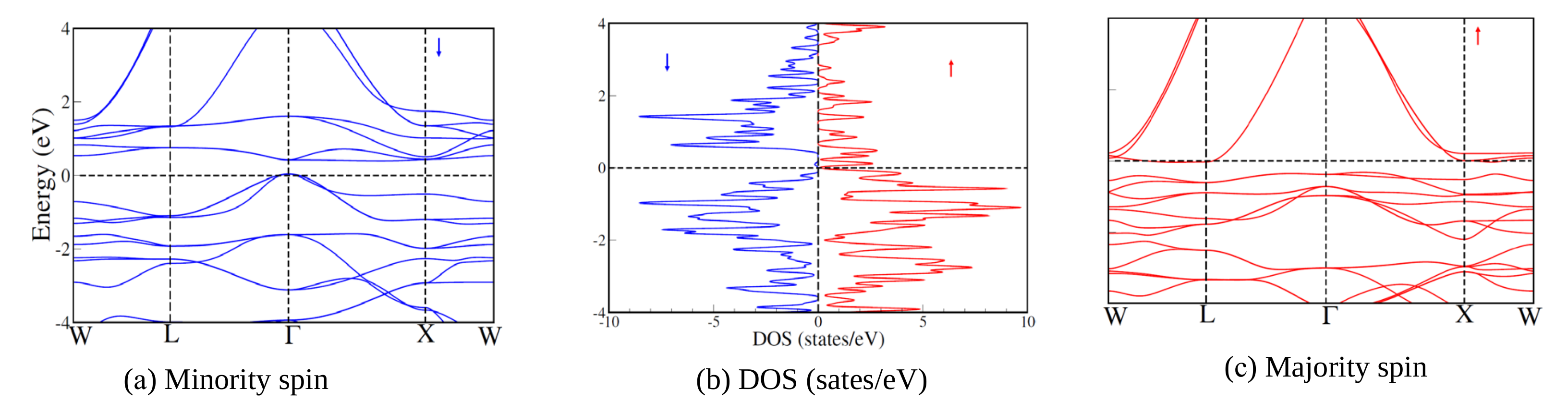}}
{\caption{Spin-polarized band structure and density of states of CoFeMnSn in Type-1 structure: (a) minority (spin-down) band (b) density of states, (c) majority (spin-up) band. The energy axis zero point has been set at the Fermi level, and the majority (spin-up) and minority (spin-down) electrons are represented by positive and negative values of the DOS, respectively.}\label{Fig_3}}
\end{figure*}

DFT studies on CoFeMnSn in the LiMgPdSn-type structure were initially carried out in order to optimize the crystal structure and find the most stable configuration. In a quaternary Heusler alloy $XX'YZ$, if the Z atoms are considered at position 4\textit{a} (0,0,0), the three remaining atoms X, $X'$, and Y might be positioned in three alternative fcc sublattices, namely 4\textit{b} (0.5,0.5,0.5), 4\textit{c} (0.25,0.25,0.25), and 4\textit{d} (0.75,0.75,0.75). Out of a total of six potential combinations, only three independent structures are viable because the permutation of the atoms in the 4\textit{c} and 4\textit{d} locations leads to energetically invariant configurations; these three independent structures are shown in Fig.~\ref{Fig_2}. Our DFT calculations indicate that Type-1 structure has minimum energy.

We have employed the GGA + U technique with Hubbard U to examine the impact of localised d electrons of transition-metal elements on the total density of states of the compound around the Fermi level. Hubbard U has been found to have very little impact on Co and Fe \textit{d} electrons and to have a significant impact only on the \textit{d} electrons of Mn ions.
Thus, we have exclusively used GGA + U on Mn d electrons to do the self-consistent computations.
The supplementary diagram~\cite{[{See Supplemental Material at }][{ for details.}]supp} illustrates the fluctuations in DOS at E$_F$ in the majority-spin state and the minimum value of total DOS near Fermi level with various U$_{eff}$ (=U$-$J, where U and J are the Coulomb and exchange parameter). Fig.~\ref{Fig_3} shows the spin-polarised band structure at U$_{eff}$ = 0.8 eV that corresponds to the lowest ground state (Type-1). In one spin sub-band, the Density of States (DOS) exhibits a band gap of 0.38 eV (semiconducting behavior), whereas, in the other spin sub-band, the Fermi level is located inside a very small energy gap. A band gap exists between the t$_{2g}$ and e$_g$ states, in the minority spin band structure. The majority-spin  band structure reveals an indirect gap where the valence band at \textit{$\Gamma$} nearly touches the conduction band at \textit{X}, which can be correlated to valley-like behavior of the DOS at \textit{E}$_F$. The majority band's closed band gap, which exhibits valley-like DOS behavior at \textit{E}$_F$, and the minority band's modest band gap (0.38 eV), both demonstrate the potential of CoFeMnSn to function as a SGS. According to a comprehensive investigation of several band-crossings around \textit{E}$_F$,
near the \textit{E}$_F$ small DOS is contributed for the majority spin band composed of d-orbital of Co and Fe. Mn remains the dominant contributor to the magnetic moments of CoFeMnSn ($\mu_{Mn}$ = 2.94 $\mu_{\rm B}$), while the contribution of  Fe and Co is less compared to Mn. The magnetic moment of Co and Fe sites are  0.70 $\mu_{\rm B}$ and 0.48 $\mu_{\rm B}$ respectively and Sn have magnetic moment -0.116 $\mu_{\rm B}$. The total magnetic moment of this system is 4.0 $\mu_{\rm B}$. Notably, it has been demonstrated that various more Ga-based Heusler alloys exhibit comparable band-crossings at the Fermi level in minority-spin states with incredibly low levels of DOS close to \textit{E}$_F$~\cite{alijani2011electronic,alijani2011quaternary}. According to predictions, these materials will be half-metallic, which means that they will be semiconducting in our case's minority-spin channel (with a negligibly tiny DOS at \textit{E}$_F$) and metallic in the majority-spin channel. In the supplementray section~\cite{[{See Supplemental Material at }][{ for details.}]supp}, we have explored the behavior of DOS near  \textit{E}$_F$ at various U$_{eff}$ (Table S1 and  Fig. S1 of supplementary material~\cite{[{See Supplemental Material at }][{ for details.}]supp}) and for various k-meshes (Fig. S2 of supplementary material~\cite{[{See Supplemental Material at }][{ for details.}]supp}).

\subsection{\label{sec:Structure}Structural analysis}
\begin{figure}[h]
\centerline{\includegraphics[width=.48\textwidth]{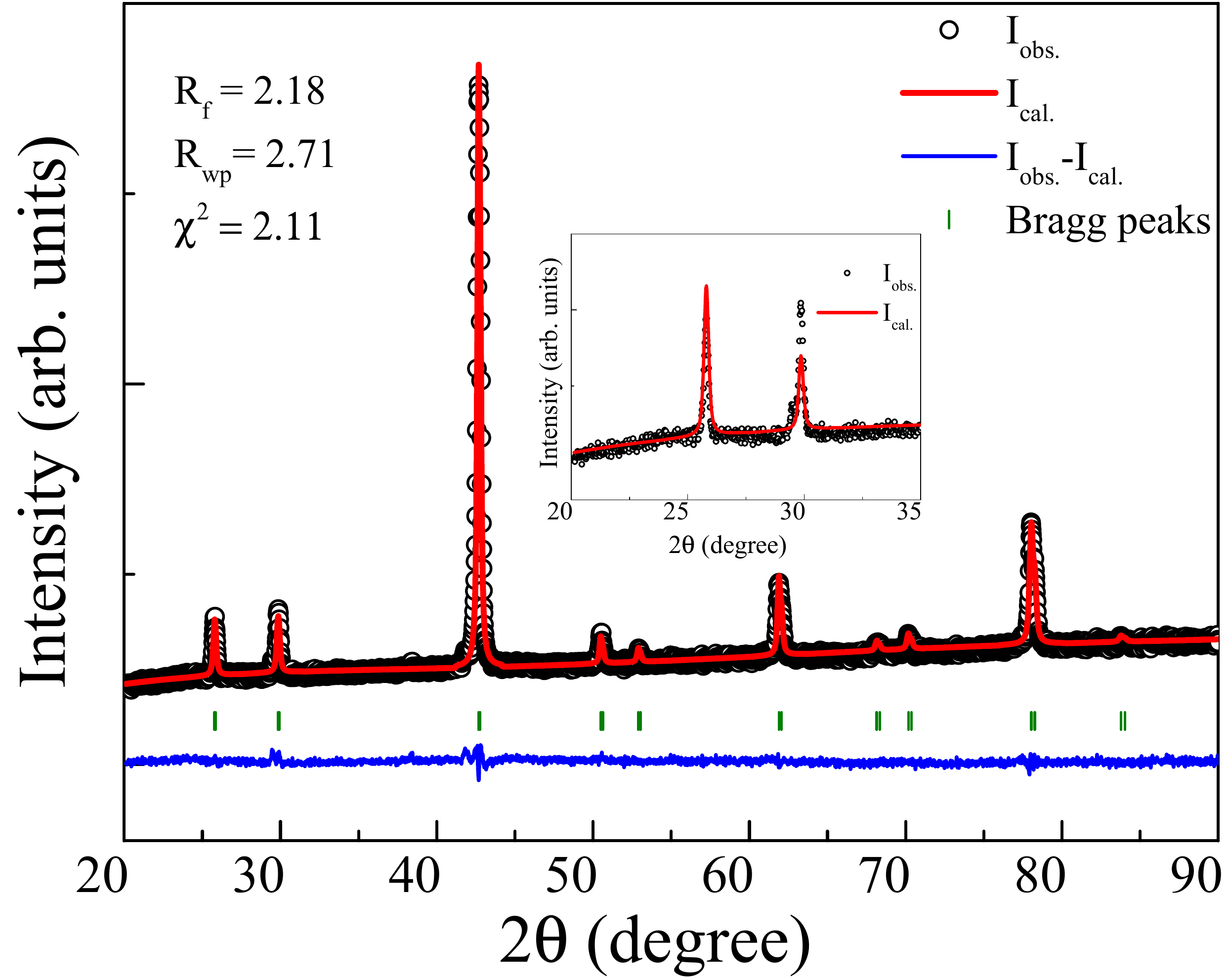}}
{\caption{Rietveld refinement of the powder X-ray diffraction pattern using off-stoichiometric composition CoFeMnSn. The Rietveld refinement data using the stoichiometric composition is shown in the inset of Fig.~\ref{Fig_4}}\label{Fig_4}}
\end{figure}

Fig.~\ref{Fig_4} represents the Rietveld refinement of the XRD pattern of CoFeMnSn taken at room temperature. As discussed above, out of three possible non-degenerate  configurations, Type-1 structure (LiMgPdSn-type) in which Sn is at 4\textit{a} (0,0,0), Mn at 4\textit{b} (0.5,0.5,0.5), Fe at 4\textit{c} (0.25,0.25,0.25) and Co at 4\textit{d} (0.75,0.75,0.75) has minimum ground state energy. We have used this Type-1 structure during the Rietveld refinement of the XRD data (Fig.~\ref{Fig_4}). The refinement confirms that the compound crystallizes in a LiMgPdSn-type crystal structure with space group $F\bar{4}3m$ (No. 216). Since both the (111) and (200) peaks are clearly visible in the XRD data presented in Fig.~\ref{Fig_4} clearly indicating towards a structural order of the studied compound. It may be noted that Rietveld refinement using stoichiometric composition CoFeMnSn is unable to explain the (200) peak (inset of Fig.~\ref{Fig_4}). We have tried different disorder (A2, B2 and DO$_3$-type) combination to fit the experimental data but fails to get a good fit. The best fit obtained with slightly off-stoichiometric composition i.e CoFe$_{1.06}$Mn$_{1.02}$Sn in ordered Type-1 structure. It may be noted here that most of these Heusler compounds exhibiting many different exotic properties, \textit{viz.}, SGS, HMF, Tunnel magneto-resistance (TMR), \textit{etc.}~\cite{felser2007spintronics} often require the material to form in a well-ordered crystal structure, as structural disorders are known to act as a great hindrance to achieve these properties~\cite{kharel2017effect,mukadam2016quantification,felser2007spintronics}.
The estimated lattice parameter from the Rietveld refinement is \textit{a$_{exp.}$} = 5.999 \AA.\\

\begin{figure}[h]
\centerline{\includegraphics[width=.48\textwidth]{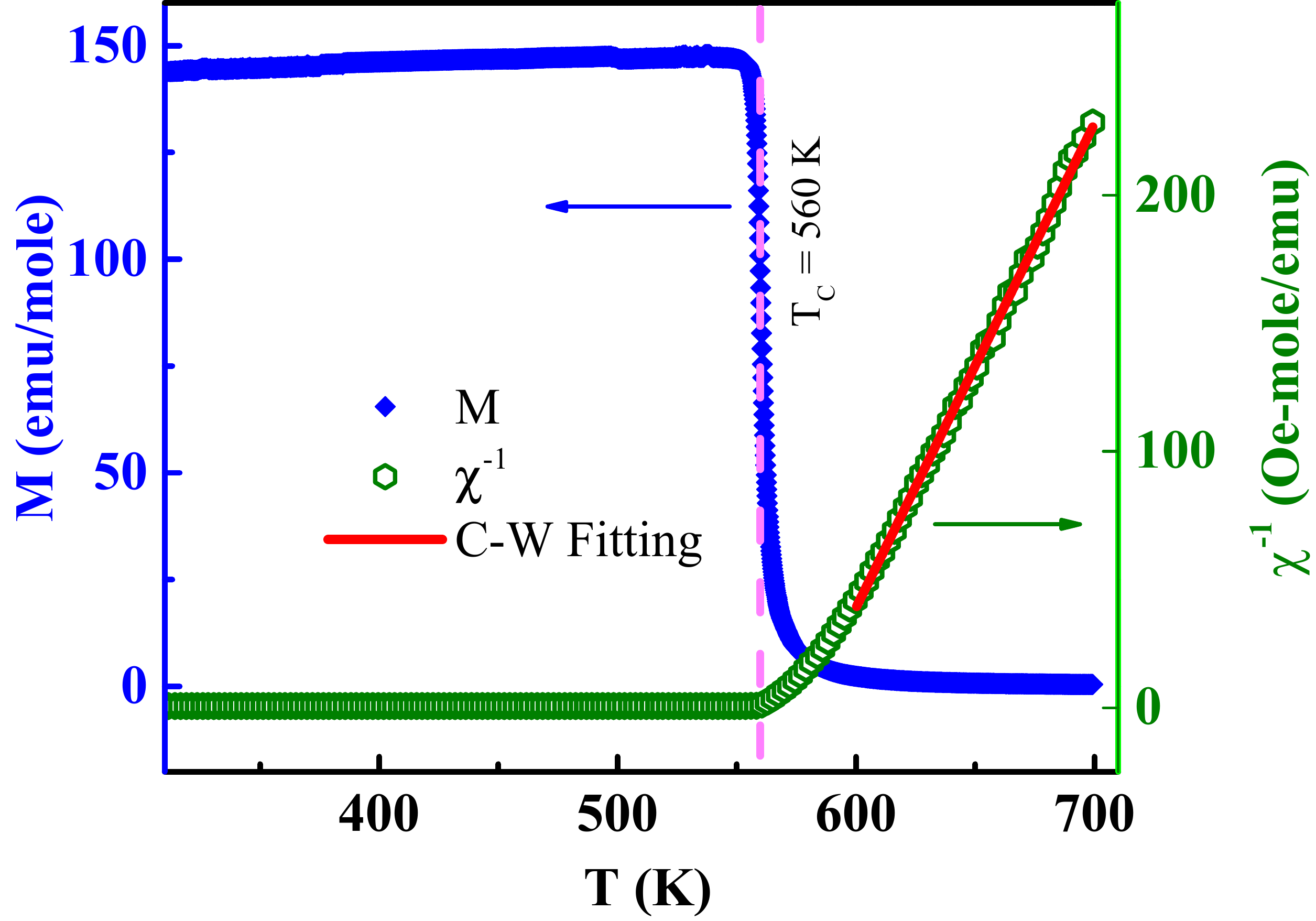}}
{\caption{ (Left Panel) Temperature dependence of magnetic susceptibility of CoFeMnSn measured in 100 Oe applied magnetic field. Curie temperature, T$_{\rm C}$, is determined from the minima in dM/dT \textit{vs.} T plot. (Right Panel) Inverse susceptibility \textit{vs.} T plot }\label{Fig_5}}
\end{figure}

\subsection{\label{sec:Magnetism}Magnetic properties}
To determine the magnetic properties of the compound, we have measured magnetization (M) \textit{vs.} temperature (T) under 100 Oe applied magnetic field (Fig.~\ref{Fig_5}). The studied compound shows ferromagnetic (FM) to paramagnetic transition ($T_{\rm C}$) near 560 K which is well above the room temperature making the compound suitable for room temperature magnetic application.  Above the transition temperature, the compound follows Curie-Weiss (C-W) behavior given by ${\chi= C/(T-\theta_P)}$, where \rm C is the Curie constant and $\theta_P$ is paramagnetic Curie-Weiss temperature~\cite{kundu2021complex,chakraborty2022ground}. C-W fit of the inverse susceptibility data in the temperature region 600--700 K yields $\mu_{eff.}$ = 4.22 $\mu_{\rm B}$/f.u\@ and $\theta_{CW}$ = 580 K which is close ($T_{\rm C}$).

\begin{figure}[h]
\centerline{\includegraphics[width=.48\textwidth]{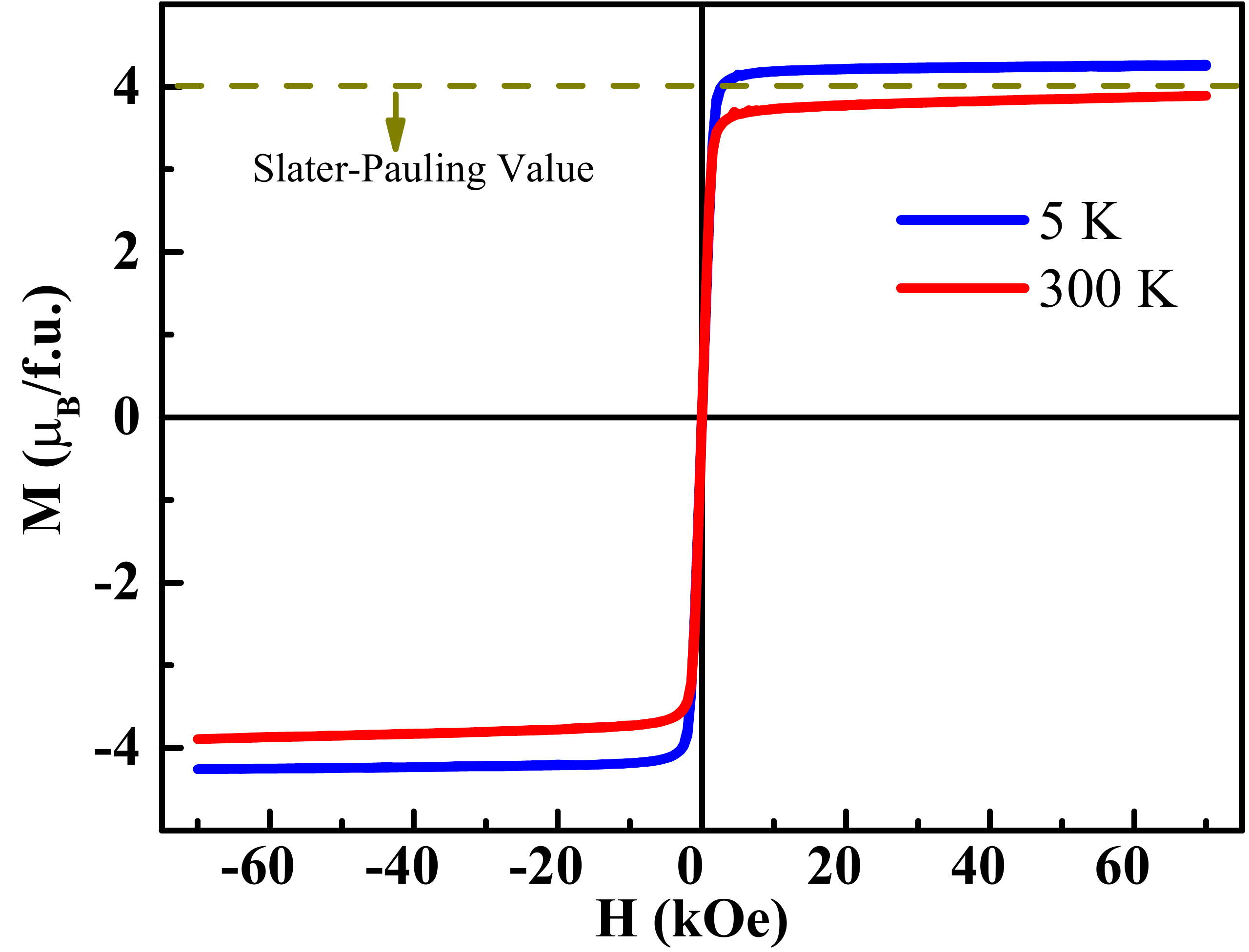}}
{\caption{Isothermal magnetization of CoFeMnSn measured at 5 K and 300 K\@.}\label{Fig_6}}
\end{figure}

The presence of the SGS nature, as predicted by DFT calculations presented above in Sec.~\ref{sec:DOS1} can be experimentally checked by studying the applicability of Slater-Pauling (S-P)~\cite{galanakis2002slater,ozdougan2013slater} rule. All the known SGS compounds are known to obey this rule. S-P rule dictates that the total magnetic moment for a Heusler alloy is \textit{m} = (N$_V$--24) $\mu_{\rm B}$/f.u, where \textit{m } is the total magnetic moment and N$_V$ is the total valence electron count (VEC) of the compound. For transition metal elements, VEC is the outer \textit{s}+\textit{d} electron count whereas for the \textit{s-p}-group elements, it is the total number of outer \textit{s}+\textit{p} electrons~\cite{bainsla2016equiatomic}. Thus, the total VEC for CoFeMnSn is 28, and the expected total magnetic moment following the S-P rule should come out to be 4 $\mu_{\rm B}$/f.u. The saturation magnetic moment estimated from the isothermal magnetization of the studied compound measured at 5 K (Fig.~\ref{Fig_6}) turned out to be 4.1 $\mu_{\rm B}$/f.u. which is consistent with the S-P rule as well as the DFT calculations. It may also be noted that such small deviation of the saturation magnetization between theory and experiment could be due to small compositional variations and also earlier observed for different SGS and HMF-based Heusler alloys~\cite{bainsla2016equiatomic}. The saturation magnetization decays only to 3.7 $\mu_{\rm B}$/f.u. at 300 K. The system exhibits a very weak hysteresis (H$_C$ $\sim$ 30 Oe), suggesting soft FM characteristics.\\

\subsection{\label{sec:Trasnport} Electrical resistivity, magnetoresistance and carrier concentration}

\begin{figure}[h]
\centerline{\includegraphics[width=.48\textwidth]{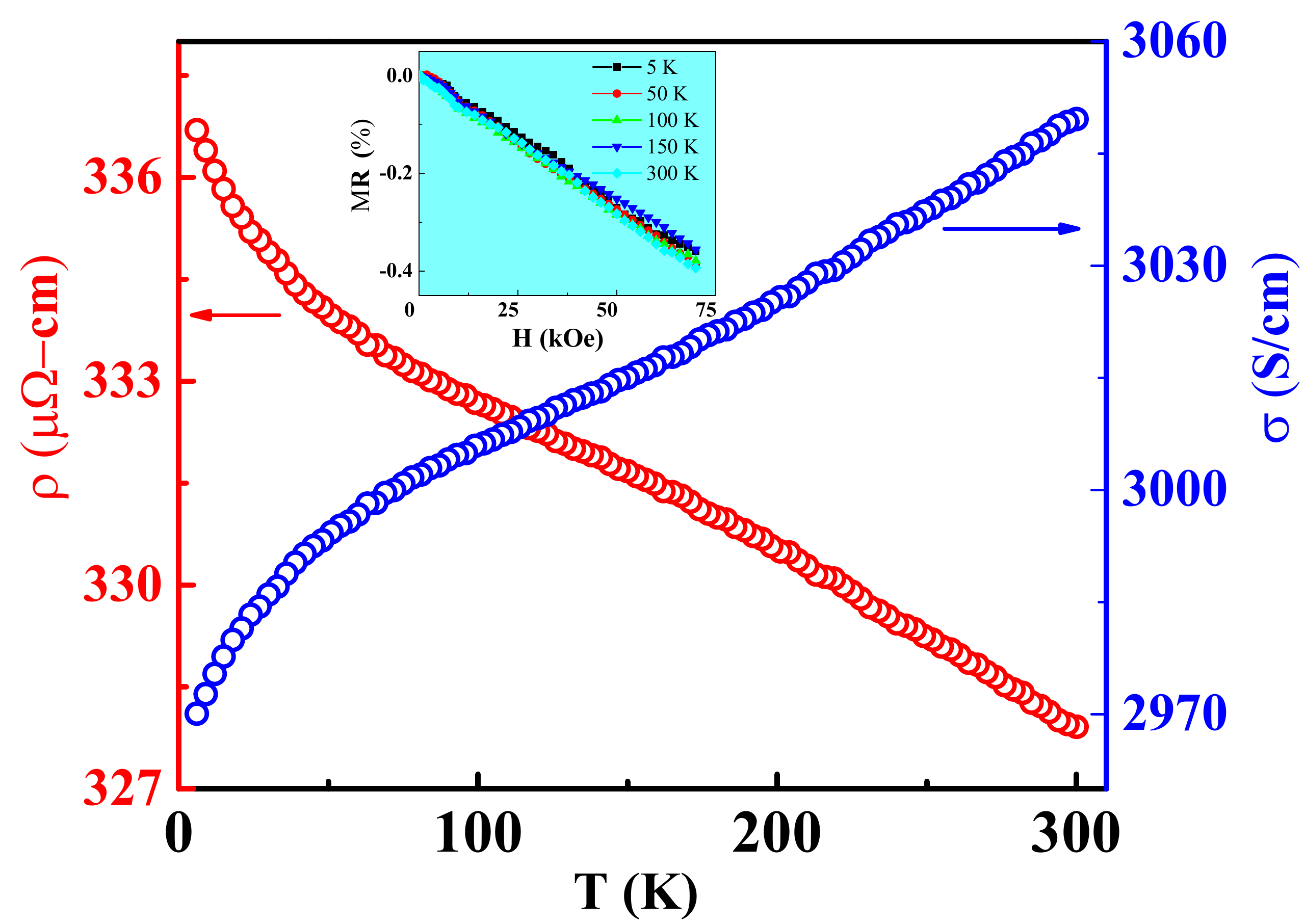}}
{\caption{Temperature dependence of the (Left-panel) electrical resistivity (Right-panel) conductivity measured in the absence of magnetic field in the temperature range 5$-$300 K. Inset shows magnetoresistance (MR) at different temperature}\label{Fig_7}}
\end{figure}

SGS characteristics are also known to cast its signatory imprint in the transport properties as well, revealing a nearly temperature-independent electrical resistivity/conductivity behavior. The temperature dependence of $\rho$(T) shows a non-metallic behaviour (Fig.~\ref{Fig_7}). The  $\rho$(T) values change from only 336 $\mu\Omega$-cm at 5 K to 327  $\mu\Omega$-cm at 300 K. The minor temperature dependence of the  $\rho$(T) confirms that this behavior could not be ascribed to the activated behavior of resistivity usually associated with semiconducting materials. Such temperature dependence of resistivity or electrical conductivity behavior is peculiar and distinctly different from that of typical metals or semiconductors but found in SGS ~\cite{ouardi2013realization,bainsla2015spin,bainsla2015origin} as well as spin-semi metallic compounds~\cite{nishino1997semiconductorlike,kuo2016ru,mondal2018ferromagnetically,venkateswara2019coexistence}. The conductivity reaches to a value of 3049 S/cm  at 300 K which is in the same range of that reported in other SGS Heusler alloy ( Mn$_2$CoAl: 2440 S/cm~\cite{ouardi2013realization}, CoFeMnSi :2980 S/sm~\cite{bainsla2015spin} and CoFeCrGa : 3233 S/cm~\cite{bainsla2015origin} and CrVTiAl : 3900 S/cm~\cite{venkateswara2018competing}), and lower than that observed for the only known spin-semimetal FeRhCrGe (4600 S/cm)~\cite{venkateswara2019coexistence}. All these together suggest the presence of SGS character in the studied compound, CoFeMnSn as well. SGS compounds are also known to show nearly non-saturating and nearly linear magneto-resistance (MR) even in the high field due to its gap-less semi-conducting nature~\cite{abrikosov1998quantum,ouardi2013realization}. To further validate the SGSs signature in the MR, we have measured MR at different temperatures in the applied magnetic field 70 kOe (inset of Fig.~\ref{Fig_7}). The magneto-resistance is found to be negligible which is of the order of $\sim$ 0.3$-$0.4 \% under the magnetic field 70 kOe at 5 K. Although the MR value is quite low, the non-saturating and linear nature of the MR at all the measured temperature range (inset of Fig.~\ref{Fig_7}) nevertheless confirms the SGS nature of the studied compound.

\begin{figure}[h]
\centerline{\includegraphics[width=.48\textwidth]{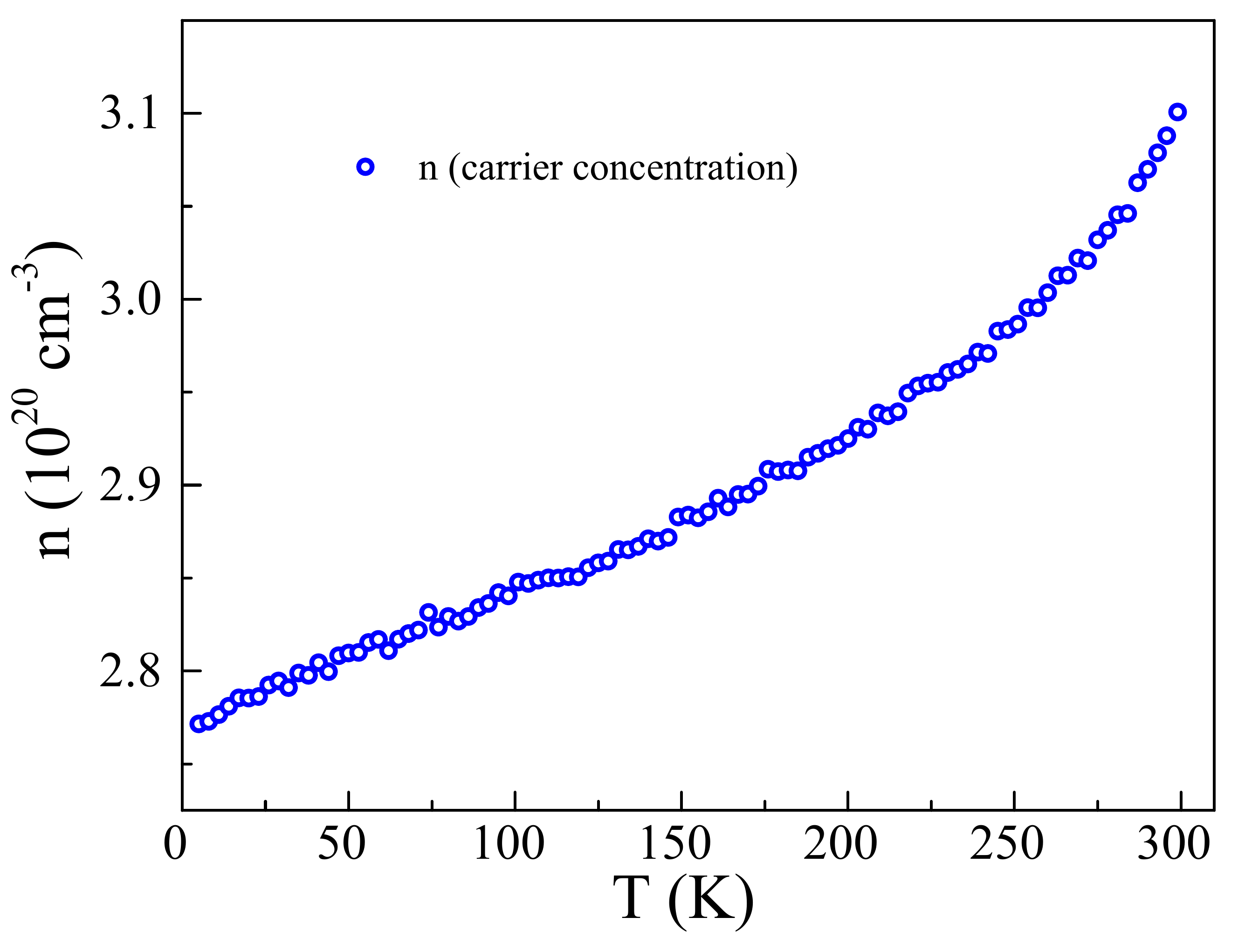}}
{\caption{Temperature dependence of the carrier concentration (\textit{n}(T)) in the temperature range 5$-$300 K.}\label{Fig_8}}
\end{figure}

\begin{figure*}[t]
\centerline{\includegraphics[width=.98\textwidth]{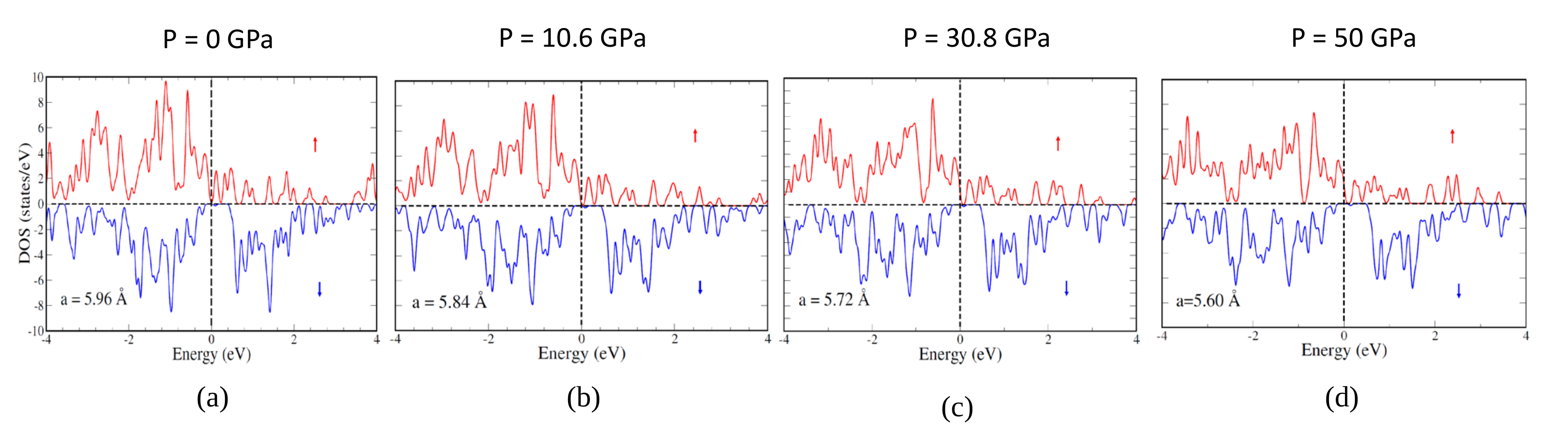}}
{\caption{Density of states at different lattice parameters (a) 5.96 \AA  (b) 5.84 \AA (c) 5.72 \AA (d) 5.60 \AA. The energy axis zero point has been set at the Fermi level, and the majority (spin-up) and minority (spin-down) electrons are represented by positive and negative values of the DOS, respectively.}\label{Fig_9}}
\end{figure*}

\begin{figure}[t]
\centerline{\includegraphics[width=.48\textwidth]{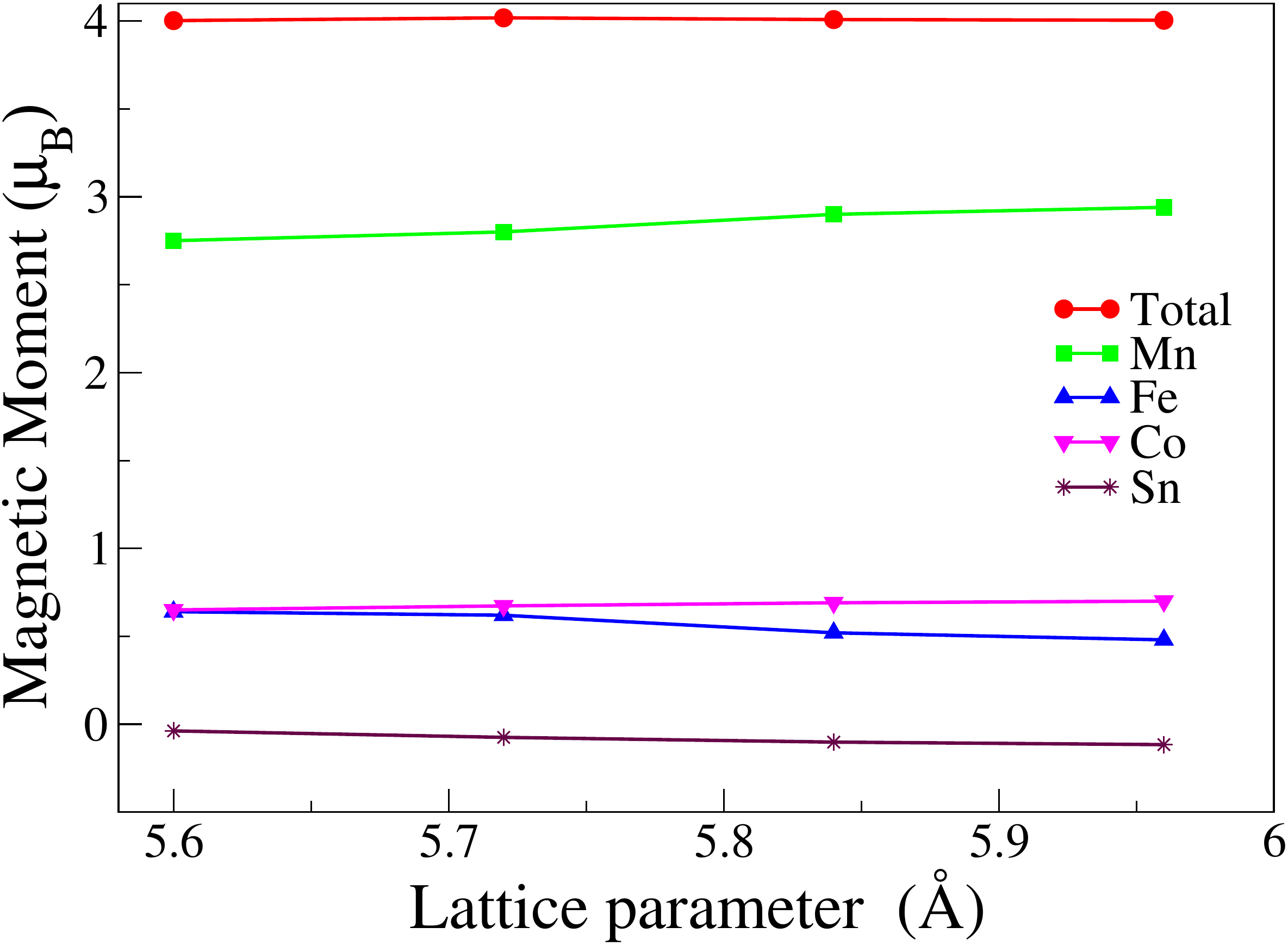}}
{\caption{ Total and site magnetic moments for different lattice parameters.}\label{Fig_10}}
\end{figure}

Another method to check the SGS characteristics is to study the temperature-dependent carrier (\textit{n(T)}) concentration of the material using the Hall measurements. Typical semiconducting materials show the exponential temperature dependence of carrier concentration, whereas SGS in contrast to semiconductors shows temperature-independent carrier concentration~\cite{bainsla2015spin}.  \textit{n}(T) remains fairly constant in the measured temperature region 5--300 K as can be clearly seen in Fig.~\ref{Fig_8}. n(T) reaches a value of 3.1$\times$10$^{20}$ cm$^{-3}$ at room temperature, which is similar to other reported bulk SGS compounds (Mn$_2$CoAl: 10$^{20}$ cm$^{-3}$~\cite{ouardi2013realization}, CoFeMnSi: 4$\times$10$^{20}$ cm$^{-3}$~\cite{bainsla2015spin} and CoFeCrGa: 10$^{20}$ cm$^{-3}$~\cite{bainsla2015origin}).

\subsection{\label{sec:Robustness}Robustness of SGS with pressure (switching from SGS to HMF )}


\begin{figure*}[t]
\centering
\includegraphics[width=1\linewidth]{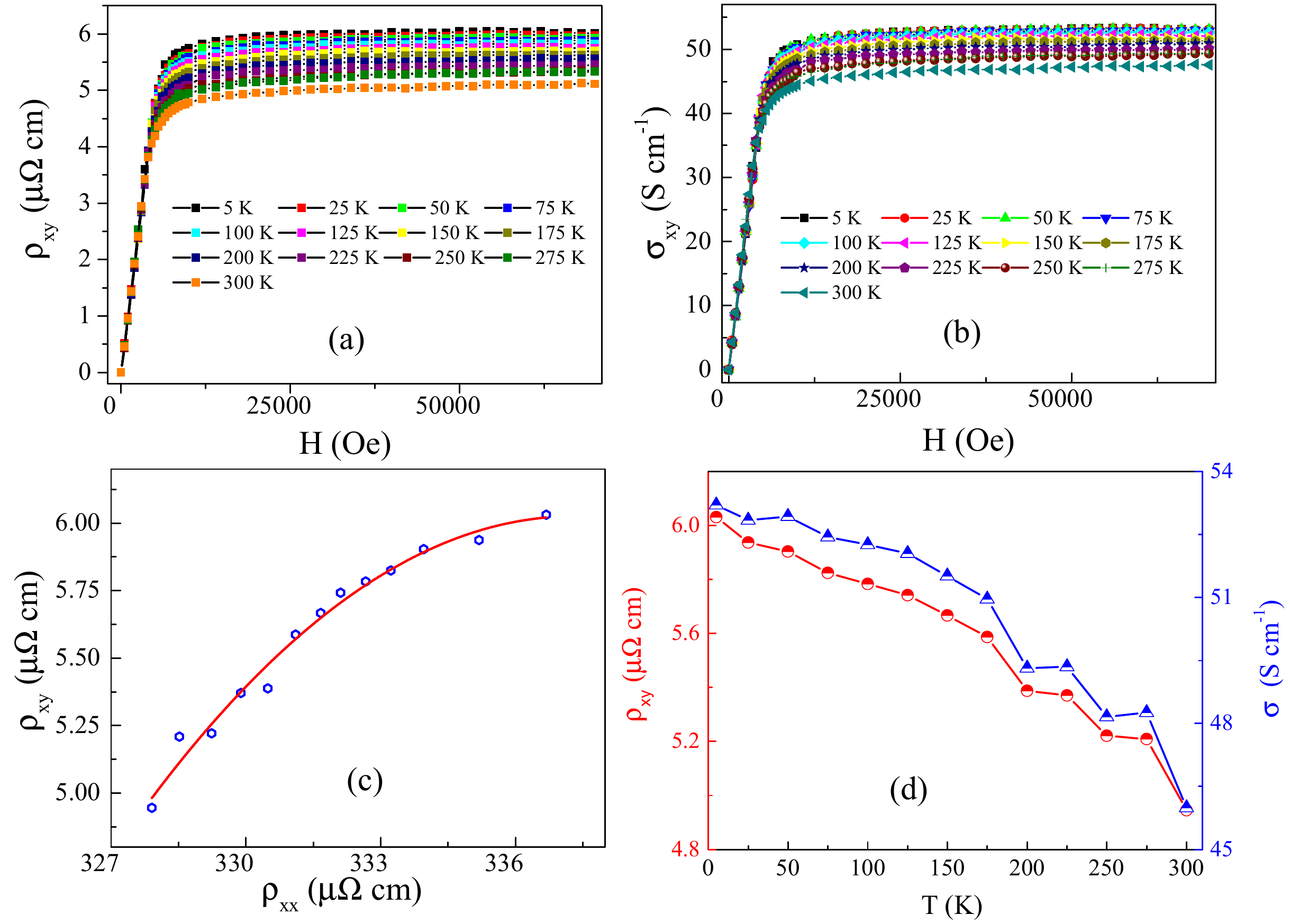}
\caption{ (a) Hall resistivity $\rho_{xy}$ (b) Hall conductivity $\sigma_{xy}$ as a function of magnetic field at different temperatures (c) plot between $\rho_{xy}$ and $\rho_{xx}$ and scaling analysis (red-line) (d) Temperature dependence $\rho_{xy}$ and $\sigma_{xy}$.}
\label{Fig_11}
\end{figure*}

\begin{figure*}[t]
\centering
\includegraphics[width=1\linewidth]{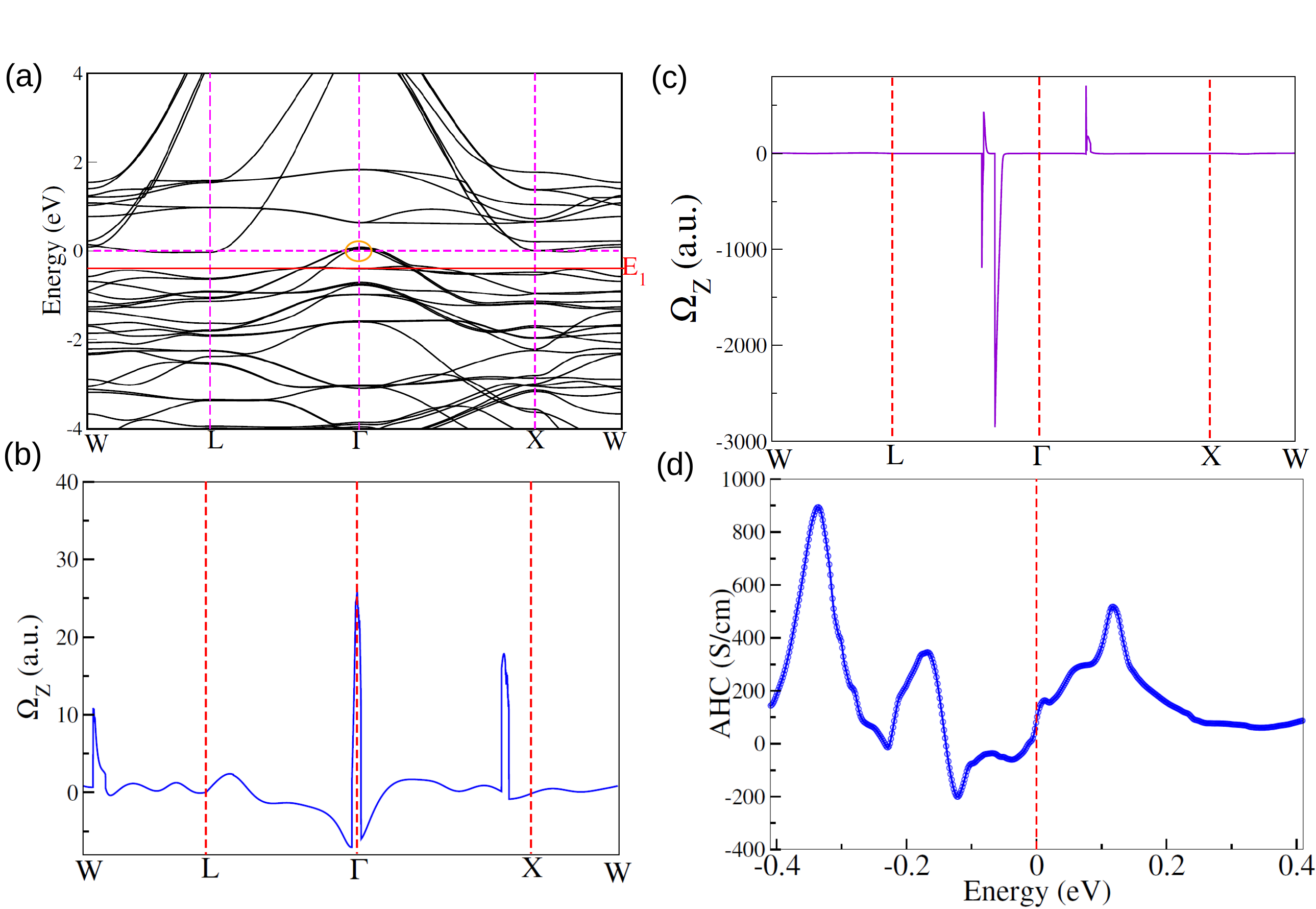}
\caption{(a) Band structure of CoFeMnSn with spin-orbit coupling (SOC), band splitting region shown by an orange circle at \textit{$\Gamma$} point. We also display the hole-doped solid red line (E$_1$ energy surface). (b) Berry Curvature corresponding to the nontrivial band at E$_F$ (c) Large negative Berry Curvature along the high symmetry point at the E$_1$ energy surface   (d) Energy (E-E$_F$ ) dependence of the AHC For CoFeMnSn.}
\label{Fig_12}
\end{figure*}

The SGS properties of the Heusler compound are sensitive to internal and external factors, such as atomic disorder, pressure, magnetic field, electric field, etc. Therefore we study the effect of pressure (P) on the electronic properties of the system especially the DOS and the applied pressure is hydro-static in nature. The DOS of CoFeMnSn for four different pressure P = 0 GPa, 10.6 GPa, 30.8 GPa, and 50 GPa are shown in Fig.~\ref{Fig_9}. We noticed that the DOS of the minority band remains gapped, whereas the majority DOS near E$_F$ is increases. However, the SGS property remains intact below lattice parameter a = 5.84 \AA(P = 10.6 GPa) and then this system shows to have finite DOS at E$_F$ which is the signature of the Half-metallic ferromagnet (HMF). It’s interesting to note that the behavior of CoFeMnSn changes from SGS to HMF when the lattice parameter is reduced or pressure is increased. The magnetic property of this system is robust to P and the total magnetic moment is almost constant with P. However there is a small variation in the magnetic moment in constituent atoms. The magnetic moment of Mn and Co decreases with pressure which goes from 2.94 $\mu_{\rm B}$ to 2.75 $\mu_{\rm B}$ for Mn and 0.70 $\mu_{\rm B}$ to 0.65 $\mu_{\rm B}$ for Co  on applying pressure P = 0 to 50 GPa, where Fe gains the magnetic moment 0.48 $\mu_{\rm B}$ to 0.64 $\mu_{\rm B}$ on applying pressure P = 0 to 50 GPa, and when pressure increased from 0 Gpa to 50 Gpa, the magnetic moment of Sn increased from -0.116 $\mu_{\rm B}$ to -0.038 $\mu_{\rm B}$ (details of the Badder charge analysis for different pressure values is provided in the supplementary material~\cite{supp}). This also shows that the magnetic behavior follows the S.P. rule. According to Fig.~\ref{Fig_10}, the total magnetic moment is essentially constant and is consistent with prior findings for different SGS ~\cite{bainsla2015origin} and HMF systems~\cite{kanomata2010magnetic}.

\subsection{\label{sec:AHC} Anomalous hall}

In comparison to HMFs and ferromagnetic metal~\cite{yao2004first}, SGSs are reported to show a lower AHC ($\sigma_{xy}$ = $\rho_{xy}$/$\rho_{xx}^2$)~\cite{ouardi2013realization,bainsla2015spin,bainsla2015origin} considering that both have same valence electron count and magnetization due to their unique band structure. For instance, the AHC was predicted to be 22 S/cm for Mn$_2$CoAl~\cite{ouardi2013realization}, 162 S/cm for CoFeMnSi~\cite{bainsla2015spin}, and a little bit higher (185 S/cm) for CoFeCrGa~\cite{bainsla2015origin}. These values are considerably less than the AHC that has been recorded for other HMF-based Heusler alloys~\cite{li2020giant,belopolski2019discovery}. The electronic structure of SGSs can be used to understand the cause of this phenomenon. SGSs compound possess a unique band structure in which the minority spin channel shows a finite gap while the majority spin channel shows a closed gap. According to reports, the minority spin states in the SGS compounds cancel out the Berry curvature of the majority spin states, resulting in a modest or nearly negligible intrinsic AHC. The half-metallic systems have a larger number of majority spin states and a finite minority spin channel gap as compared to SGS compounds, which leads to a larger intrinsic AHC~\cite{manna2018colossal,shahi2022anti}.
In order to verify the SGS characteristics of CoFeMnSn, the hall conductivity has been measured at different temperatures. The $\sigma_{xy}$  at 5 K is presented in the Fig.~\ref{Fig_11} (b) and it mimics the nature of isothermal magnetization. \\

Total AHC ($\sigma_{xy}$) is estimated to be 53 S/cm which is consistent with other reported SGSs system~\cite{ouardi2013realization,bainsla2015spin,bainsla2015origin} and validates the presence of SGS state in the studied compound. According to general agreement, there are two primary mechanisms that affect the anomalous Hall effect (AHE): an intrinsic mechanism connected to the band structure's Berry curvature and an extrinsic mechanism connected to impurity scattering. Skew dispersion and side-jump are two subclasses of the extrinsic mechanism. Skew scattering is exactly proportional to ${\rho_{xx}}$ and results from unequal scattering brought on by impurity scattering. Contrarily, a side-jump is a lateral jump brought on by impurity dispersion. It can be difficult to distinguish between side-jump and the intrinsic mechanism because they are both proportional to ${\rho_{xx}}$$^2$. The scaling relation~\cite{tian2009proper} describing anomalous Hall can be written as
$\rho$${_{xy}^{A}}$ = a${\rho_0}$+b(${\rho_{xx}-\rho_0}$)+c${\rho_{xx}^{2}}$, where ${\rho_0}$ is the residual resistivity, a and b represent defect and phonon induced skew scattering, respecpetively. The parameter c indicates the combined impact of the Berry curvature and side-jump effects. The fitted parameters are found to be a = 3.9, b = 7.9 and c = -0.01, from the fitting it can be understood that both extrinsic and intrinsic mechanism contribute to the AHE. The negative value of coefficients (a,b) signifies that extrinsic skew scattering contribution is in opposite  direction to both side-jump and intrinsic contribution due to momentum space Berry curvature, resulting a lower total AHC which is almost equal to half of the theoretically predicted intrinsic contribution. Similar features have been also observed for other SGSs~\cite{rani2019spin,shahi2022anti}, FeRhCrGe~\cite{venkateswara2019coexistence} and MnGa~\cite{mendoncca2017spin}.

\subsection{\label{sec:Theory2} Anomalous Transport behaviour}

 To understand the intrinsic contribution to transport properties the electronic band dispersion is studied in presence of SOC.  The band splitting at \textit{E}$_F$ is noticed after the introduction of SOC  when we take into account magnetization in the [001] direction, as illustrated in Fig. \ref{Fig_12}(a). The band splitting~\cite{shukla2022band,wang2006ab} region is encircled at $\Gamma$-point as shown in   Fig. \ref{Fig_12}(a). We also calculate Berry-curvature (BC) along high symmetry points shown in Fig. \ref{Fig_12}(b). We estimated the Z component of BC at the \textit{E}$_F$ energy surface, and the result is quite low.

The electronic motion exhibits a transverse anomalous velocity due to the Berry curvature, which results in a significant AHC. The intrinsic AHC can be assessed using the Kubo formalism's linear response theory, and it can be represented as 

\begin{equation}
 \sigma_{xy} = {\frac{e^2}{\hbar} \int\frac{d^{3}\textit{k}}{(2\pi)^3}\sum_{n}\Omega^z_{n}(\textit{k})f_n(\textit{k})}  \end{equation}
 $\Omega^{z}_n$  is Berry curvature and it can be written as
 \begin{eqnarray}
\Omega^z_{n} = -2i \sum_{m \neq n} \frac{{\langle \psi_{n\textit{k}}|v_x|\psi_{m\textit{k}}\rangle} {\langle \psi_{m\textit{k}}|v_y|\psi_{n\textit{k}}\rangle}} {[E_{m}(\textit{k}) - E_{n}(\textit{k})]^2}
\end{eqnarray}
\noindent
where $f_{n}(\textit{k})$ is the Fermi-Dirac distribution function, \textit{n} is index of the occupied bands, $E_{n}(\textit{k})$ is the eigenvalue of the n$^{th}$ eigenstate $\psi_{n}(\textit{k})$, $v_i$ = $\frac{1}{\hbar}\frac{\partial H(\textit{k})}{\partial \textit{k}_i }$ is the velocity operator along the \textit{i (i = x, y, z)} direction. The spin-orbit coupling is taken into account along [001], which is thought of as the direction of magnetic polarisation, to calculate AHC. Theoretically predicted intrinsic AHC (86 S/cm) is in good agreement with experimentally observed total AHC (53 S/cm). To understand the dependence of chemical potential shift or hole/electron doping we have done energy-dependent of AHC as shown in Fig.~\ref{Fig_12} (d). The AHC increases on electron doping whereas on doping  hole initially, the AHC decreases up to -0.16 eV and further increases upto -0.2 eV. On the other hand, the value of BC turns out to be quite high along the \textit{$\Gamma$}-\textit{L} high symmetry point at the E$_1$ energy surface (-0.33 eV, where the AHC has greatest predicted AHC, See Fig.~\ref{Fig_12}(c)). Additionally, there are limited tiny BC contributions in other directions. A single electron can be removed from the primitive cell to achieve this energy surface (E$_1$).

\section{Conclusion}
In conclusion, we report a novel SGS compound, CoFeMnSn, in the equiatomic quaternary Heusler alloy family through different theoretical calculations and experimental investigations. Reitveld refinement of the XRD data confirms the ordered structure of the studied compound in which Sn occupies 4\textit{a} position, Mn at 4\textit{b} position, Fe at 4\textit{c }position and Co at 4\textit{d} position consistent with the structure optimization calculation. Spin-polarized band structure calculations in the most stable structure reveal the SGS nature of the compound. The high transition temperature (T$_{\rm C}$ = 560 K) of the studied compound makes the material suitable for room temperature applications. The saturation moment estimated from the isothermal magnetization measurements at different temperatures in the range of 5-300 K was found to be nearly invariant (4.1--3.7) $\mu_B$/f.u. and obeys the S-P rule which is a prerequisite to the SGC characteristics. This result is also consistent with the DFT calculation for this material. The compound's nearly temperature-independent conductivity, resistance and carrier concentration in the temperature range of 5-300 K  along with low AHC at 5 K further support the SGS nature. The robustness of the SGS property has been also examined theoretically. Theoretical simulations also show that hole doping may be used to get a high intrinsic AHC value in the examined material. The combined SGS and topological features of CoFeMnSn thus make it a potential candidate for future spintronic devices.\\


\section{Acknowledgement}
S.G.  would like to sincerely acknowledge SINP, India for the Ph.D. fellowship.  J.S. thanks the University Grants Commission (UGC) for support through a Ph.D. fellowship.
\bibliographystyle{apsrev4-2}
\normalem
%

\end{document}